\documentclass[a4paper,fleqn,usenatbib]{mnras}

\usepackage{newtxtext,newtxmath}
\usepackage[T1]{fontenc}
\usepackage{ae,aecompl}

\usepackage{graphicx}	% Including figure files
\usepackage{amsmath}	% Advanced maths commands
\usepackage{amssymb}	% Extra maths symbols
\usepackage{float}
\usepackage{subfig}
\usepackage{enumitem}
\newlength\mylen
\setlist[itemize,1]{leftmargin=*,labelsep=-\mylen}

\newcommand{\hi}{H\,{\sc{i}}\,}

\newcommand{\msol}{M$_{\odot}$}
\newcommand{\kms}{km\,s$^{-1}$}

\title{Enhanced \hi profile asymmetries in close galaxy pairs  }

\author[Bok et al.]{
J. Bok,$^{1,2}$\thanks{E-mail: jamie@ast.uct.ac.za}
S.-L. Blyth,$^{2}$
D.G. Gilbank$^{1,3}$
E.C. Elson$^{4,5}$
\\
$^{1}$South African Astronomical Observatory, Observatory, Cape Town, 7935,  South Africa \\
$^{2}$Department of Astronomy, University of Cape Town, Private Bag X3, Rondebosch 7700, South Africa \\
$^{3}$Centre for Space Research, North-West University, Potchefstroom 2520 Cape Town, South Africa \\
$^{4}$Department of Physics $\&$ Astronomy, University of the Western Cape, Robert Sobukwe Rd, Bellville, 7535, South Africa \\
$^{5}$South African Radio Astronomy Observatory (SARAO), Observatory, 7925, South Africa }

\date{Accepted XXX. Received YYY; in original form ZZZ}

\pubyear{2015}

\hypersetup{draft}
\begin{document}

\label{firstpage}
\pagerange{\pageref{firstpage}--\pageref{lastpage}}
\maketitle
% * <jamie@ast.uct.ac.za> 2016-12-08T18:44:18.918Z:
%
% ^.
% * <jamie@ast.uct.ac.za> 2016-11-24T07:01:46.962Z:
\begin{abstract}
Analyzing the quantified \hi profile asymmetries of galaxies in different environments, we explore not only the prevalence of asymmetry in \hi profiles, but also the possibility of using \hi profile asymmetries to trace merger activity. We construct close pair and isolated galaxy catalogues of \hi profiles from the Arecibo Legacy Fast ALFA (ALFALFA) survey, and using a simple \hi flux ratio, quantify and compare the profile asymmetries between the two catalogues. In this way we investigate the popular proposition that merger activity causes \hi profiles to become asymmetric, and thereby probe the role of mergers in galaxy evolution. We find small but significant differences between the asymmetry distributions of the two samples, indicating that merger activity does indeed enhance asymmetry in the global \hi profile.   
\end{abstract}


%%%%%%%%%%%%%%%%%%%%%%%%%%%%%%%%%%%%%%%%%%%%%%%%%%

%%%%%%%%%%%%%%%%% BODY OF PAPER %%%%%%%%%%%%%%%%%%

\section{Introduction}

\noindent Numerous observations have shown galaxies to exist in a vast variety of shapes and sizes, colours and kinematics, interaction states and environments. Astronomers have endeavored to decipher the mechanisms and physical processes by which such variety in galaxy properties arises, and significant advances in our understanding of how galaxies form and evolve have been made. According to the theoretical model for structure formation, Lambda CDM, galaxies form and evolve hierarchically through a succession of major and/or minor mergers \citep{WhiteRees1978}. Observationally and theoretically, galaxy mergers (major and minor) have been pointed to as key processes in various aspects of galaxy formation and evolution \citep{Mundy2017}. The proposition that elliptical galaxies might be the product of a major merger between spirals, the so-called `merger hypothesis' was first proposed forty years ago by \cite{Toomre1977}. Subsequent observations of merger remnants and faint shells/tidal features around ellipticals provide convincing support for the picture of galaxy evolution via mergers \citep{Hopkins2009}.  \cite{FernandezLorenzo2013} demonstrate the importance of environment on the growth in size of massive spirals, which they show to have larger sizes than samples of less isolated galaxies.  Mergers have also been implicated in the observed growth in stellar mass of massive galaxies by a factor of 2-3 from $z\sim$~$2$ to the present, as well as the observed increase in size by a factor of 3-6 of massive quiescent galaxies at fixed stellar mass in the same time period (see \cite{Mundy2017} and references therein).  Observing galaxies in the process of merging therefore presents a unique opportunity to test galaxy evolution models, and thereby enable the development of a more complete picture of galaxy evolution.

The general strategy for observing mergers in the optical regime is based on the fact that the gravitational interplay between merging galaxies can cause the interacting galaxies to become morphologically disturbed. Late stage mergers are thus identified as having highly disturbed morphologies, while merging galaxies are often characterized by the presence of tidal features. Close pairs of galaxies (merger candidates) on the other hand, mark the beginning stages of the merger process. The states of non-equilibrium induced by merger activity in both the stellar and dark matter components of a galaxy translate into asymmetries in the galaxy stellar light distribution \citep{Reichard2007}, and it is these asymmetries that have become useful tracers of merger activity in the optical regime. Previously asymmetries in optical images have been quantified using  Fourier decomposition \citep{Reichard2007,Jog2009}, CAS parameters \citep{Conselice2003}, and other 2D methods \citep{Schade1995}. 
%the $m=1$ azimuthal Fourier mode \citep{Reichard2007,Jog2009}, the CAS parameters (concentration, asymmetry, clumpiness) \citep{Conselice2003}, and the $R_T + R_A$ index \citep{Schade1995}. 
Such techniques have been used to show that galaxies in close pairs exhibit enhanced asymmetries in their stellar light distributions compared to galaxies that are isolated \mbox{\citep{Patton2005,DePropis2007,Ellison2010}}.
%The techniques mentioned above were developed on optical data, however, 
\cite{Jog2009}, and references therein, showed that galaxies are asymmetric not only in their stellar populations, but also in their gas (molecular and neutral) distributions, kinematics (see \cite{Swaters1999}, \cite{Califa2015}), and global \hi velocity (spectral) profiles. It was \cite{Baldwin1980} who coined the term 'lopsided' in 1980, reserving the title for galaxies in which they detected an asymmetry in the spatial extent of their neutral gas in their pioneering paper `Lopsided galaxies'.  Comparing asymmetries traced in the optical with asymmetries traced by \hi, as in the work of \cite{Kornreich2000}, showed that asymmetry is not only quantitatively larger and more frequent in \hi than in stars \citep{Bournaud:2005jt}, but also that the amplitude of asymmetry increases with galaxy radius \citep{Reichard2007}. This, together with the fact that \hi typically extends out to much larger radii than the stellar component of a galaxy, suggests that \hi might be a more sensitive probe of asymmetry compared to the optical light distribution. 

Looking to \hi imaging as a potential diagnostic for tracing asymmetries associated with merger activity, \cite{Holwerda2011} quantified the \hi morphologies of a sample of 141 column density maps of galaxies from the WHISP survey, computing the CAS parameters for each galaxy as per \cite{Conselice2003}, as well as $M_{20}$, the Gini coefficient \citep{Lotz2004}, and $G_M$ (the second order moment of light). The results of the study suggest that disturbed morphologies and asymmetries are indeed good indicators of merger activity. Following up on the work of \cite{Holwerda2011}, \cite{Giese2016} investigated the dependence of these morphological parameters on signal-to-noise ratio, resolution, and inclination, and also found that the asymmetry parameter is the most useful parameter with which to measure galaxy lopsidedness as traced by classifications by eye. \cite{Califa2015} show kinematic asymmetries/misalignments in both the spatially resolved stellar and ionised gas components of galaxies are also good indicators of interaction status. Combining N-body/hydro-dynamics/stellar evolution code, \cite{Kornreich2002} simulated the dynamics and morphology of a galaxy in response to receiving a librational `kick' of energy, and found that NGC 5474 (a tidally disturbed galaxy) exhibited almost all of the observed effects in the simulation. \cite{Kornreich2002} suggest that driven sloshing modes might play a role in galaxy asymmetry.

While the 2D imaging analyses are very promising, our current \hi imaging datasets have limited statistics. However, large \hi surveys such as HIPASS and ALFALFA consisting of thousands of spatially unresolved, but spectrally resolved detections, can be investigated. These surveys were conducted using single-dish radio telescopes, which do not spatially resolve galaxies with sizes smaller than the primary beam, however, they do provide spectrally resolved information (global \hi velocity profiles) for large numbers of galaxies.  

The 1D global \hi profile has a shape primarily dictated by galaxy kinematics, and carries with it information both about a galaxy's velocity field, as well as the distribution of \hi within the galaxy.  The ordered motions within a galaxy, where radial velocities tend to cluster, are responsible for the characteristic double-horn signature seen in \hi velocity profiles. \hi profiles are therefore expected to be symmetric about the systemic velocity for an unperturbed disk, with deviations from symmetry considered potential consequences of merger activity (e.g. non-circular motions, tidal tails, and distortions in the \hi mass distribution), asymmetric gas accretion, or an offset of the stellar disc in a halo potential \citep{Eymeren2008}. Furthermore, \cite{RichterSancisi1994} found that \hi profile asymmetries are often accompanied by an asymmetry in the corresponding \hi gas distributions. 

The freqency of these observed departures from symmetry has been the topic of investigation in a number of \hi profile studies, where the degree of asymmetry has been assessed both qualitatively and quantitatively \citep{Haynes1998, Matthews1998, Espada2011}.  \cite{RichterSancisi1994} qualitatively measured the \hi profile asymmetry on a sample of $\sim1700$~  galaxies from various single dish \hi surveys using a by eye visual classification scheme. This sample was comprised primarily of field galaxies so as to minimize the chance of the cluster environment playing a role in producing profile asymmetries. A lower limit of 50 percent of the sample was found to exhibit significant profile asymmetries, suggesting that asymmetry might well be the rule rather than the exception. \cite{RichterSancisi1994} further quantified the \hi profile asymmetries of their sample using \cite{Tifft1990}'s \hi flux ratio between the lower and upper velocity halves of the global \hi spectrum, and found the qualitative and quantitative asymmetry measures to be highly congruent. Using a similar \hi flux ratio, \cite{Haynes1998} quantified the \hi profile asymmetry for an isolated sample of 104  high signal-to-noise (SNR) \hi profiles obtained using the Greenbank 43m telescope. 
%Using the Arecibo General Catalogue (AGC) to search for companions, care was taken to select only those galaxies that are \hi isolated out to 1 square degree, and with no companions within 400 \kms to avoid the asymmetry results being contaminated by artificial asymmetries produced by confusion.
The results of the study showed $\sim50$~ percent of the sample to have significant \hi profile asymmetries (in good agreement with previous work), which the authors attribute to distortions in the \hi distribution, non-circular motions, and possible confusion with unidentified companions within the telescope beam. A more recent study by \cite{Espada2011}, specifically focussed on \hi profile asymmetries of galaxies carefully selected to be isolated. Their study forms part of the AMIGA project (Analysis of the interstellar Medium in Isolated GAlaxies \cite{VerdesMontenegro2005}), whose aim is to disentangle those galaxy properties (morphological and structural) which are due to internal secular evolution from those which arise from interactions within the galaxy environment.  \hi profile asymmetries were quantified for a sample of $\sim$166 high SNR \hi profiles (the \hi refined sub-sample) using the standard \hi flux ratio. They describe the resulting asymmetry distribution as following a Gaussian model with width $\sigma = 0.13$ (corresponding to a flux ratio of 1.26 at the 2$\sigma$ level).

While previous studies show \hi profile asymmetries to be a common phenomenon, their origin is still unclear.  Since merger activity is known to induce asymmetries in the 2D \hi distributions of galaxies, we propose that it produces asymmetries in \hi profiles as well. Other potential drivers of asymmetry include harassment (\cite{Moore1996}), ram-pressure stripping (\cite{Gunn1972}), viscous stripping (\cite{Nulsen1982}), outflows (\cite{Fraternali2017}), and accretion (e.g. \cite{Sancisi2008}), and while in this paper we investigate mergers in particular as an asymmetry driver, we note that it is difficult to isolate drivers without conducting a detailed environment study. Here we explore the relationship between asymmetries in \hi profiles and possible merger activity by quantifying the profile asymmetries of a sample of close galaxy pairs (merger candidates), and comparing them with the \hi profile asymmetries of a reference sample of isolated galaxies. By investigating first the extreme case of close pair galaxies, where we expect the signal in \hi profile asymmetry to be strongest, we explore the possibility of using \hi profile asymmetries as a way to identify merger activity on different scales (loose groups, dense groups, clusters) in the future, in lieu of what is currently a very limited sample of \hi maps. Upcoming surveys on SKA pathfinders will build up a more complete set of HI maps out to intermediate redshifts, $~$0.2-0.3, but at higher redshifts the resolution will generally be too poor to provide an accurate measure of asymmetry (see Fig, 9 from \cite{Giese2016}) and therefore the 1D \hi profile will be very useful as a consistent means to measure asymmetry over a wide range of redshifts.

This paper is organized as follows: In the following section we discuss the different data sets used in this study, and in section \ref{sampleselection} we describe our sample selection criteria for both the pair and isolated galaxy samples. Section \ref{method} outlines our method for quantifying \hi profile asymmetry, including a description of how we estimate uncertainty. We discuss results and future work in section \ref{results}, and summarize the conclusions in section \ref{summary}. Throughout this paper we adopt  $H_0=70 $ \kms Mpc$^{-1}$ ($h = 1$), $\Omega_{M}= 0.3$, and $\Omega_{\Lambda} = 0.7$.

%Here we investigate the relationship between asymmetries in \hi profiles and possible merger activity by quantifying the profile asymmetries of a sample of close galaxy pairs (merger candidates), and comparing them with the \hi profile asymmetries of a reference sample of isolated galaxies. In this way we explore the possibility of using \hi profile asymmetries as a way to identify merger activity, and thereby probe the role of mergers in galaxy evolution. We use \hi profiles from the Arecibo Legacy Fast ALFA (ALFALFA) blind \hi survey of the nearby Universe ($-1600$ \kms $\leq z \leq 18 000$ \kms), and a spectroscopic catalogue of SDSS DR7 (Sloan Digital Sky Survey \cite{Abazajian2009} ) galaxies within the ALFALFA footprint to assess the degree of isolation of each \hi galaxy (close-pair vs. isolated). 

\section{Data}\label{data}

%such as Section~\ref{sec:maths} below. \\ \\ 
The close pair and isolated galaxy catalogues we wish to construct require both a sample of \hi galaxy profiles, as well as an optical sample of galaxies from which we can draw optical neighbours for each \hi galaxy. We require not only positional information, but reliable redshift information such that we can compute the 2D projected distance between nearest neighbours.

\subsection{HI Galaxy sample}
\noindent This study uses publicly available \hi profiles from the first data release of the Arecibo Legacy Fast ALFA (ALFALFA) survey \citep{Giovanelli2005}, $\alpha 40$. The $\alpha 40$ catalogue \citep{Haynes2011} covers $40\%$ (2800~square degrees) of the total survey area. Source centroid positions, \hi line flux densities, recessional velocities and line widths are provided for a total of 15855 sources in the catalogue, as well as the most probable optical counterparts (OCs) in SDSS DR7 (Sloan Digital Sky Survey Data Release 7 \citep{Abazajian2009}) for more than $98$ percent of the \hi sources. OCs were identified by the $\alpha 40$ team using distance from the \hi centroid, as well as colour, morphology, and redshift information that was publicly available at the time. 

We make use of a sub-sample of high signal-to-noise, good quality \hi profiles (reliable detections flagged by the $\alpha 40$ team as code 1 profiles) with spectroscopic optical counterparts. A pool of $\sim$6800 \hi galaxies meet these selection criteria. Our \hi sample has a velocity resolution of 5 \kms out to z $\sim 0.06$.%Arecibo's relatively small beam size ($\sim3.5'$) and large collecting area probes the faint end of the \hi mass function in the local universe. 

\subsection{Optical neighbours}
Neighbours for our \hi sample, which we use to investigate the environment of each \hi galaxy, are drawn from a spectroscopic sample of SDSS galaxies that are within both the ALFALFA footprint and redshift range. 
The SDSS galaxy sample is made up of galaxies with r-band Petrosian magnitudes r $\leq$ 17.77 and r-band Petrosian half-light surface brightnesses $\mu_{50}$ $\leq$ 24.5 mag arcse$c^{-2}$, above which the sample is $>99$ percent complete \citep{Lupton2002}. 
\cite{Lupton2002} show that redshifts for the SDSS galaxy sample are reliable with a statistical error less than 30 \kms, and \cite{Toribio2011} show that the dispersion
in the difference between radial velocities of ALFALFA galaxies and their assigned optical counterparts in SDSS is $\sim$35 \kms.

\section{Sample selection} \label{sampleselection}
In order to compare the quantified \hi profile asymmetries of galaxies in close pairs with those that are isolated, we first need to compile a galaxy pair  catalogue of \hi profiles. Here we look to previous work to inform our definition of close pair galaxies.

\subsection{Merger pair sample}\label{secpairs}
In deciding upon a useful pair definition with which to identify galaxy pairs, one must engage in a compromise between purity of the sample, and completeness. While a stringent pair definition is preferable in selecting pairs that are most likely going to merge in a relatively short timescale ($\sim$~few Gyr) (purity) , this may lead to a statistically insignificant sample size if the corresponding survey is not sufficiently large (completeness).   \cite{Barnes1988} and \cite{Patton1996} estimate that galaxies with companions at projected separations of $r_p\leq 20 h^{-1}$ kpc will merge within $~0.5$ Gyr, and the convention of early close pair studies was to use this projected separation as an upper limit for identifying merger pairs. More recently, with spectroscopic redshift samples increasing in size, it has been possible to include a velocity separation criterion, $\Delta v$, in the galaxy pair definition \citep{Patton2000}. By determining the line-of-sight rest frame velocity difference between companion galaxies one can identify those pairs with the greatest likelihood of merging as having the lowest relative velocities. 

\cite{Patton2000} use a visual interaction classification parameter based on optical morphologies to investigate the location of interacting pairs in $r_p-\Delta v$ phase space, and find that the majority of pairs showing clear signs of interactions (tidal tails, morphological distortions/asymmetries, etc.) have projected separations of $r_p\leq 20 h^{-1}$ kpc. Furthermore, $97\%$ of the pairs with $\Delta v\geq 600~$ \kms display no signs of interaction, while the strongest signs of interaction are seen for the low $\Delta v$ pairs. \cite{Patton2000} thus suggest close pair criteria of $r_p^{max} \leq 20 h^{-1}$ kpc and $\Delta v^{max} \leq 500$ \kms to identify interacting pairs. \cite{Robotham2014} conduct a similar analysis on a sample of optical pairs with $r_p^{max} \leq 100 h^{-1}$ kpc and $\Delta v^{max} \leq 1000$ \kms, and find again that pairs with the smallest projected separations ($r_p^{max} \leq 20 h^{-1}$ kpc) and lowest relative velocities ($\Delta v^{max} \leq 500$ \kms) exhibit the strongest signs of interaction. These close pair criteria are commonly used in the literature, however it is important to note that they do not recover all interacting pairs. Signs of interaction are observed at $r_p \geq 50 h^{-1}$ kpc, even as far as $r_p \sim 95 h^{-1}$ kpc in the case of Arp 295a/b \citep{Patton2000}. These systems, however, are not dominant. More recently, \cite{Patton2016} find that optical asymmetries are most significantly enhanced by the presence of nearby companions at projected separations less than 10 kpc, with the mean asymmetry enhanced by a factor of $2.0\pm0.2$ in this regime. Beyond 10 kpc the mean asymmetry enhancement declines, remaining statistically significant out to projected separations of 50 kpc. \\
It is important to note that it has been optical signs of interaction advising the close pair definitions in previous work. A close pair study in which \hi is being used as an interaction diagnostic warrants careful consideration. \hi typically extends further out compared to the optical component of a galaxy, and is more diffuse in nature. This suggests signs of interaction in \hi might be observed at larger projected separations, indicating merger activity on a different timescale to that of optical indicators. For our work, we therefore use \cite{Robotham2014}'s loosest close pair criteria of $r_p \leq 100 h^{-1}$kpc and $\Delta v \leq 500$ \kms.  In this way we hope to recover as many interacting pairs as possible (and thereby obtain a more complete sample), as well as those pairs in the early stages of the merger process, the evidence of which might only be visible in \hi.\\

\begin{figure*}
\centering
  \includegraphics[scale = 0.7]{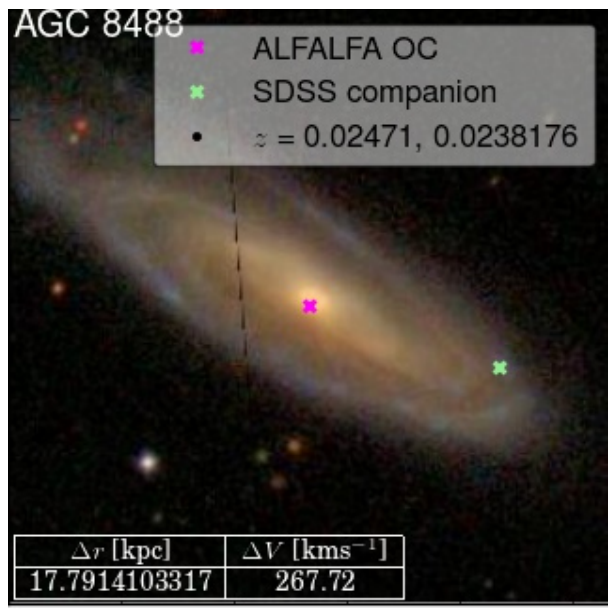}
\caption{4'x4' optical image of a  galaxy pair that was classified as potentially shredded galaxies during the visual inspection process.}
\label{shredded}
\end{figure*}

The following is a summary of the steps taken to compile our catalogue of merger-pair galaxy candidates:

\begin{enumerate}

\item Using the optical spectroscopic counterpart R.A. and Dec. positions for each \hi galaxy in the $\alpha 40$ sample with signal-to-noise (SNR) > 10 and inclination ($i$) > 30\degr, we search in the SDSS spectroscopic catalogue for the nearest (2D projected distance) neighbour galaxy. (The SNR selection criterion is necessary for the accurate measurement of \hi profile asymmetries, and is consistent with the \hi asymmetry work of \cite{Tifft1990} and \cite{Espada2011}. The inclination criterion ensures we are not underestimating the \hi profile asymmetry by including face-on galaxies in our sample, whose \hi profiles will be single peaked by virtue of their inclination with respect to our line of sight. )

\item $\Delta v = |v_{\rm{HI}} - v_{\rm{optical}}|$ is then determined for each pair using the $v_{\rm{helio}}$ from the $\alpha 40$ catalogue, and $cz$ from SDSS. We identify 375 close pair galaxies as those with at least one optical neighbour within 100~kpc and 500~\kms. 

\item In order to reduce the potential effects of confusion on our measurements of HI profile asymmetry, we removed all \hi-\hi pairs with spatial separation less than 3.5 arcminutes (size of the ALFALFA primary beam). We removed 15 close pairs using this criterion. By excluding \hi-\hi pairs we note the caveat that we are not considering all close pairs in our analysis, but a sub-sample of close pairs in which the HI content resides primarily in one galaxy.

\item A final visual inspection of the close pair galaxies eliminates 12 potentially `shredded' galaxies from the sample. As a result of the DR7 deblending process, bright objects are on occasion interpreted as two or more objects (shredded). We show an example in figure \ref{shredded}. The final pair sample thus comprises 348 galaxy pairs.

\end{enumerate}

\subsection{Sample of isolated galaxies}\label{seciso}

\begin{figure}
  \centering
  \begin{minipage}[b]{0.4\textwidth}
    \includegraphics[width=\textwidth]{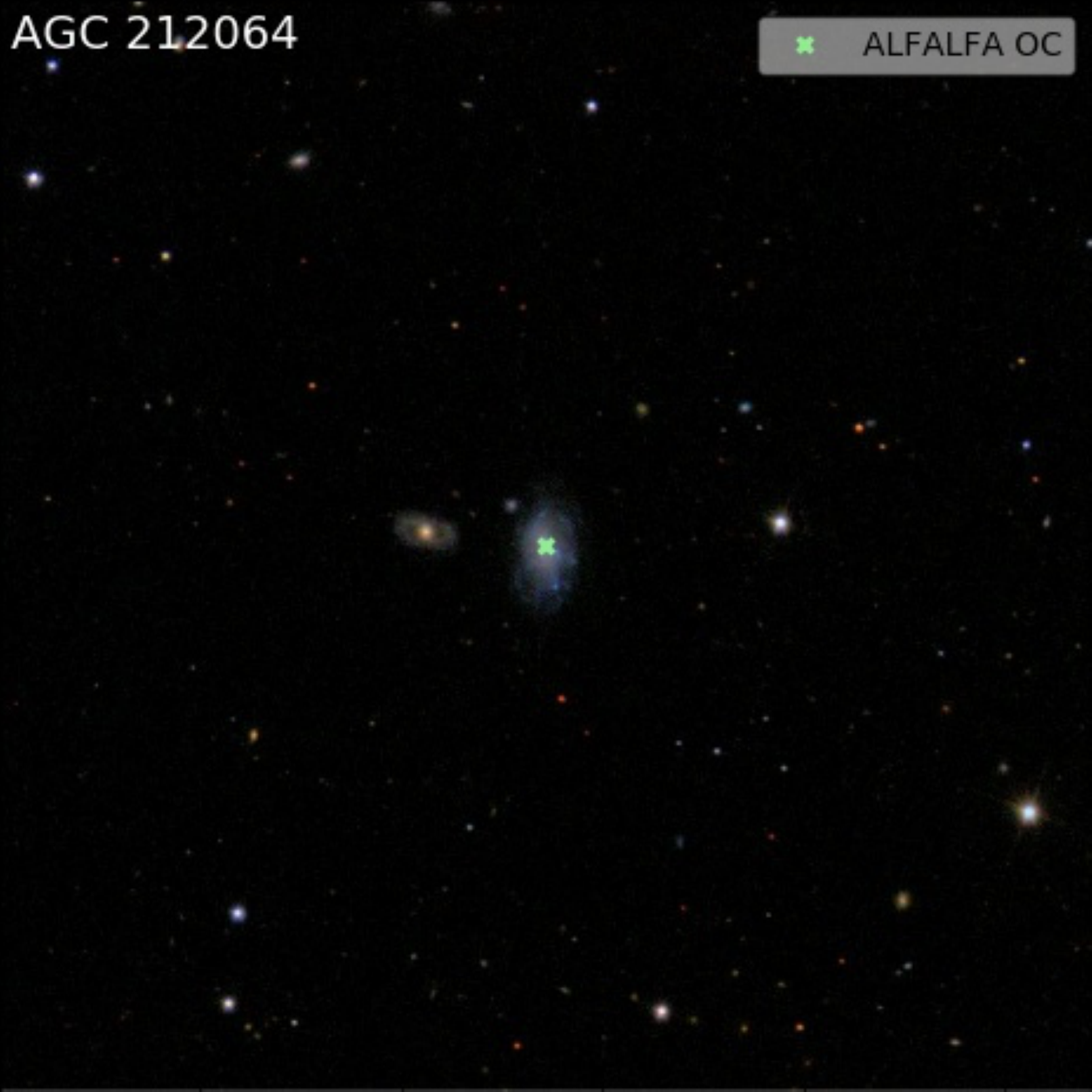}
  \end{minipage}
  \hfill
  \begin{minipage}[b]{0.4\textwidth}
    \includegraphics[width=\textwidth]{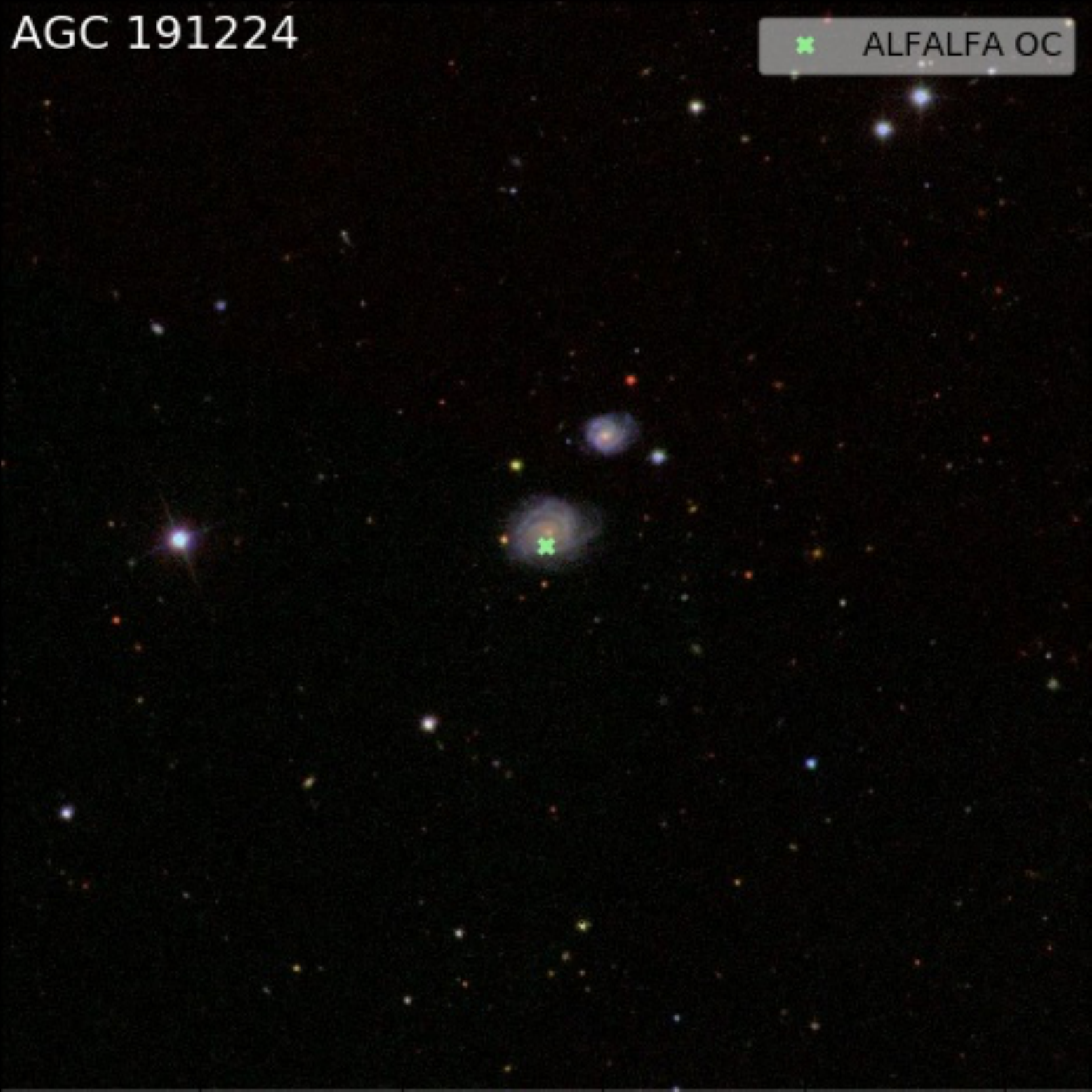}
  \end{minipage}
\caption{7'x7' ($\sim$~500 kpc) optical images of galaxies that were removed from our isolated sample due to the presence of potential neighbours. The green cross marks the location of the OC.}
\label{falseisolated}
\end{figure}

In defining the isolated galaxy sample we prioritize purity over completeness, and conservatively select only those ALFALFA galaxies whose nearest spectroscopic optical companion is $\geq 500$ kpc away, with $\Delta v \geq 5000$ \kms. At such large separations one can reasonably expect the optical companions to have negligible tidal influence on their distant \hi neighbours, and therefore are unlikely to produce asymmetries in the \hi profiles. Outside these criteria signs of optical interaction are uncommon, and increasingly insignificant \citep{Patton2000}. We note that spectroscopic incompleteness of the SDSS optical sample could affect the purity of our isolated galaxy sample, as well as the completeness of a our pair sample. If a galaxy's true nearest neighbour does not have a measured redshift in SDSS, the distance to its nearest neighbour will be over estimated, and could potentially lead to a real close pair member being classified as isolated by our isolation criteria. To this end we visually inspected 500 kpc x 500 kpc optical images of our isolated sample, and removed potential contaminants (possible pairs). In figure \ref{falseisolated} we show two such examples of galaxies that were originally classified as isolated using only spectroscopic information, but which were removed after visual inspection revealed potential companions. Our final sample of isolated galaxies comprises 304 galaxies.  

We note that while spectral profile asymmetries for an isolated galaxy sample have already been measured by \cite{Espada2011}, we propose that a control sample of our own catalogue of isolated ALFALFA galaxies, subject to the same systematics as our pair sample, will provide the most reliable comparison for our study, however we do compare to \cite{Espada2011} in section \ref{results} as well.

\subsection{Combined sample properties}\label{sampleprop}

% who suggest that in order to make a good estimate of the asymmetry in an \hi profile, the profile must have a signal-to-noise ratio greater than 10. \cite{Espada2011} adopt the same SNR criterion. 
A summary of the final selection criteria for both the pair and isolated galaxy samples can be seen in Table \ref{table:1}.

\begin{table*}
\caption{Sample selection criteria.}
\centering
%\begin{tabular}{ |p{1.1cm}||p{1cm}|p{1.2cm}|p{0.8cm}|p{1cm}  }
\begin{tabular}{lcccccc}
 \hline
Catalogue& $\Delta r$ [kpc] & $\Delta v$ [\kms] & SNR & $i$ &sample size & $z$ matched sample\\
 \hline
 \hline
 Pairs   & < 100    & < 500 &   > 10 &   >30& 348 & 304\\
 Isolated&   > 500  & > 5000   &> 10 & >30& 304 &304\\

 \hline

\end{tabular}
\label{table:1}
\end{table*}

\begin{figure*}
\includegraphics[scale=0.72]{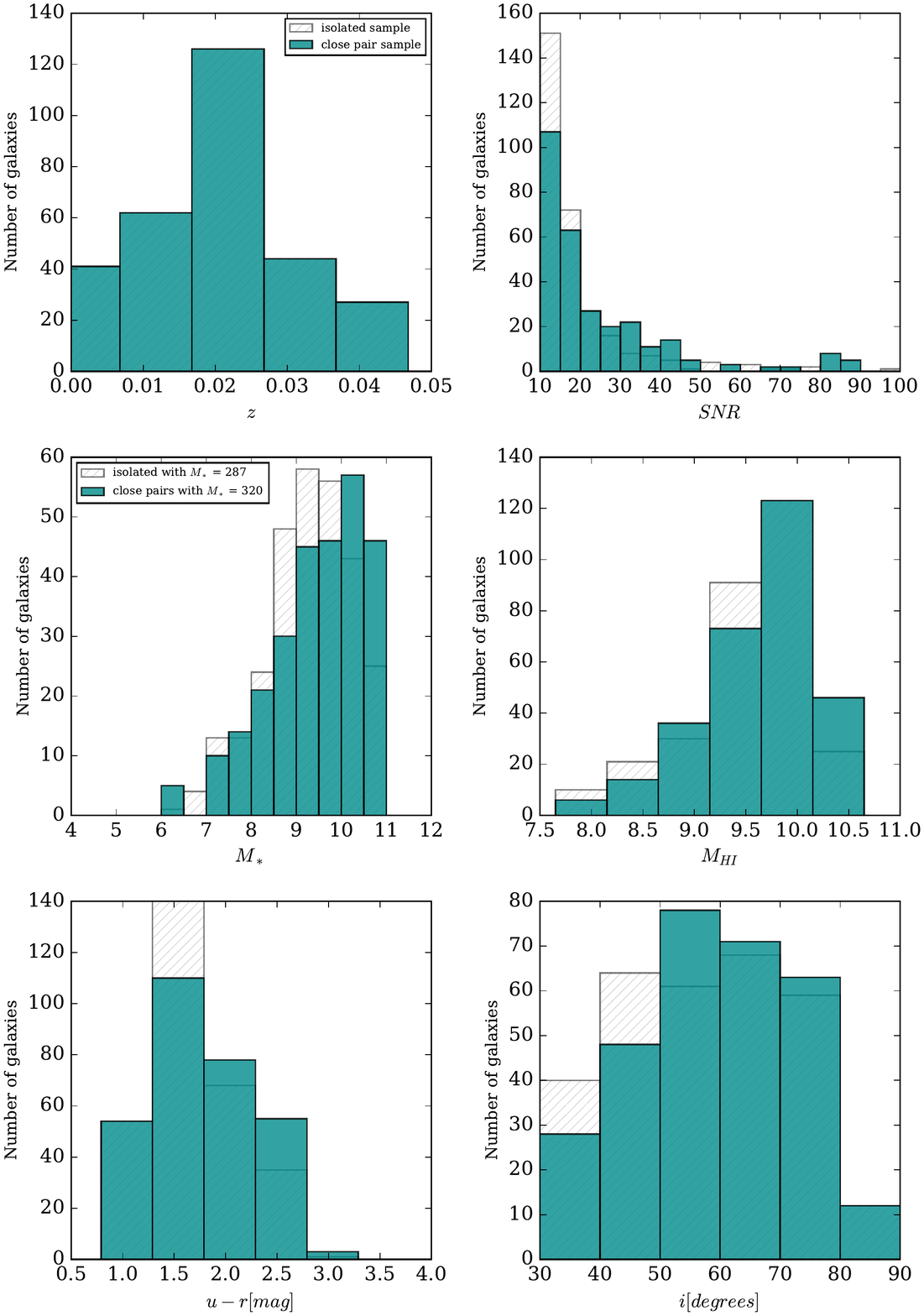}
\caption{Comparison of properties between the pair (dark cyan) and isolated (hatched) samples. Top left: redshift. Top right: SNR. Middle left: $log10(M_{* })$. Middle right: $log10(M_{HI})$. Bottom left: u-r colour. Bottom right: inclination. } 
\label{sample}
\end{figure*}

Since galaxy evolution is strongly dependent on $z$ we match the pair and isolated samples in this quantity, and compare sample properties in figure \ref{sample}. We note in figure \ref{sample} that the pair and isolated samples are well matched in SNR, and u-r colour, and thus disregard the possible impact of these quantities on the comparative measurement of \hi profile asymmetry between the two samples. Since a higher fraction of the pair sample has both larger stellar and \hi masses, as well as higher inclinations, compared to the isolated sample, we test the dependence of our asymmetry measure on these quantities in section \ref{results}.

\section{Measuring profile asymmetries}\label{method}
A simple and meaningful way to quantify asymmetries in \hi profiles is to compute an \hi flux ratio between the two profile horns, for example by using the median (/mean) velocity as the divider (e.g. \citealt{Haynes1998,Espada2011}). Variations of this method include using different quantities to determine profile edges (velocity width at the 50 percent level ($w_{50}$), velocity width at the 20 percent level ($w_{20}$)), as well as different profile centres ($v_{\rm{mean}}$, $v_{\rm{median}}$, $v_{\rm{weighted}}$). 

Here we quantify the asymmetry of our pair sample as a ratio of flux, $A_{c}$,  between the two velocity horns using $v_{helio}$ from the $\alpha40$ catalogue to define the centre of the profile. In defining the profile edges we use the  $w_{50}$ width given in the $\alpha40$ catalogue, and we interpolate the profile to retrieve the velocity at the 50 percent level. The typical flux values at $v_{20}$ are similar to the noise rms of the profiles, for this reason we use $w_{50}$ over $w_{20}$ to determine the profile edges, as it is more reliably determined. $A_c$ is calculated as:

\begin{equation}
\centering
A_{\frac{l}{h}} = \frac{\int_{v_{\rm{low}}}^{v_{\rm{helio}}}I_{\nu}dv}{\int_{v_{\rm{helio}}}^{v_{\rm{high}}}I{\nu}dv} 
\label{eq:ac}
\end{equation} \\

\noindent and 

\begin{align}
A_c&= A_{\frac{l}{h}} ~\rm{if} ~A_{\frac{l}{h}} > 1\\
&= A_{\frac{h}{l}} ~\rm{if} ~A_{\frac{l}{h}} < 1
\end{align}

\begin{figure*}
  \includegraphics[scale = 0.55]{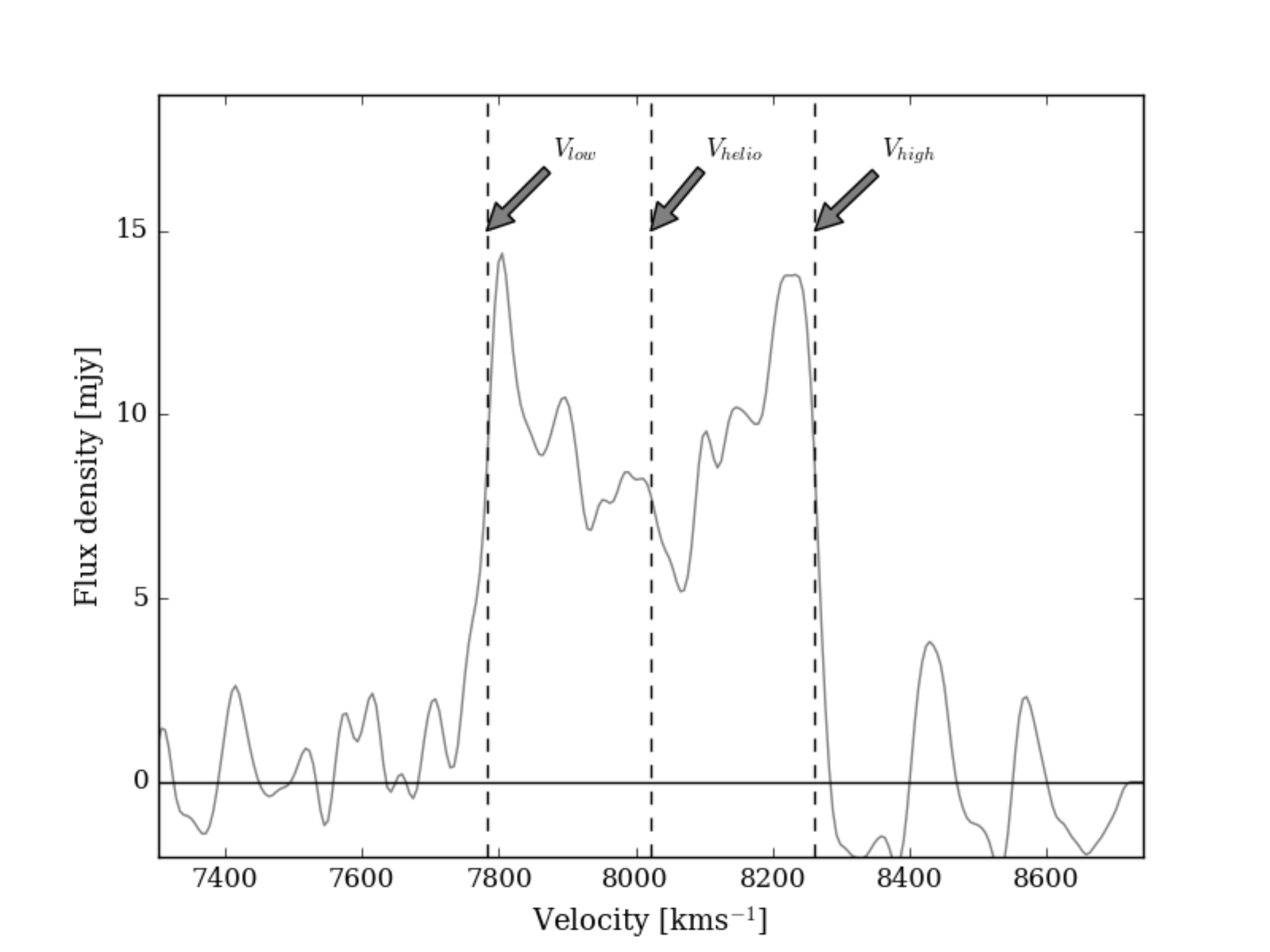}
\caption{Graphical representation of how the $A_c$ ratio is calculated on a global \hi profile. Black  vertical dashed lines mark the profile center ($V_{helio}$), as well as  the extent of the left and right velocity horns ($V_{low}$ and $V_{high}$ respectively), as taken from the ALFALFA $\alpha_{40}$ catalogue.} 
\label{ac_profile}
\end{figure*}

\noindent Here $v_{\rm{helio}}$ corresponds to the the galaxy's heliocentric velocity, and  $v_{\rm{low}}$ and $v_{\rm{high}}$ the velocity of the galaxy at the 50 percent flux level to the left and right of $v_{\rm{helio}}$ respectively. Figure \ref{ac_profile} provides a graphical representation of how $A_C$ is determined.

\subsection{$A_c$ uncertainty estimation}
We adopt a Monte-Carlo approach to estimate the uncertainty associated with the $A_c$ measurement. The steps we followed were:
\begin{enumerate}
\item{Calculate $A_c$ on the original \hi profile.}
\item{Replace each flux value, $f_i$, in the profile with a new flux value, $f_{\rm{new}}$, drawn randomly from the Gaussian distribution with mean $f_i$ and width = $\rm{rms}_{\rm{noise}}$ (as provided by the $\alpha40$ catalogue for each galaxy).}
\item{Recalculate $A_c$ on the adjusted profile.}
\item{Repeat steps (ii) and (iii) 1000 times.}
\end{enumerate}

\begin{figure*}  
\includegraphics[scale = 0.5]{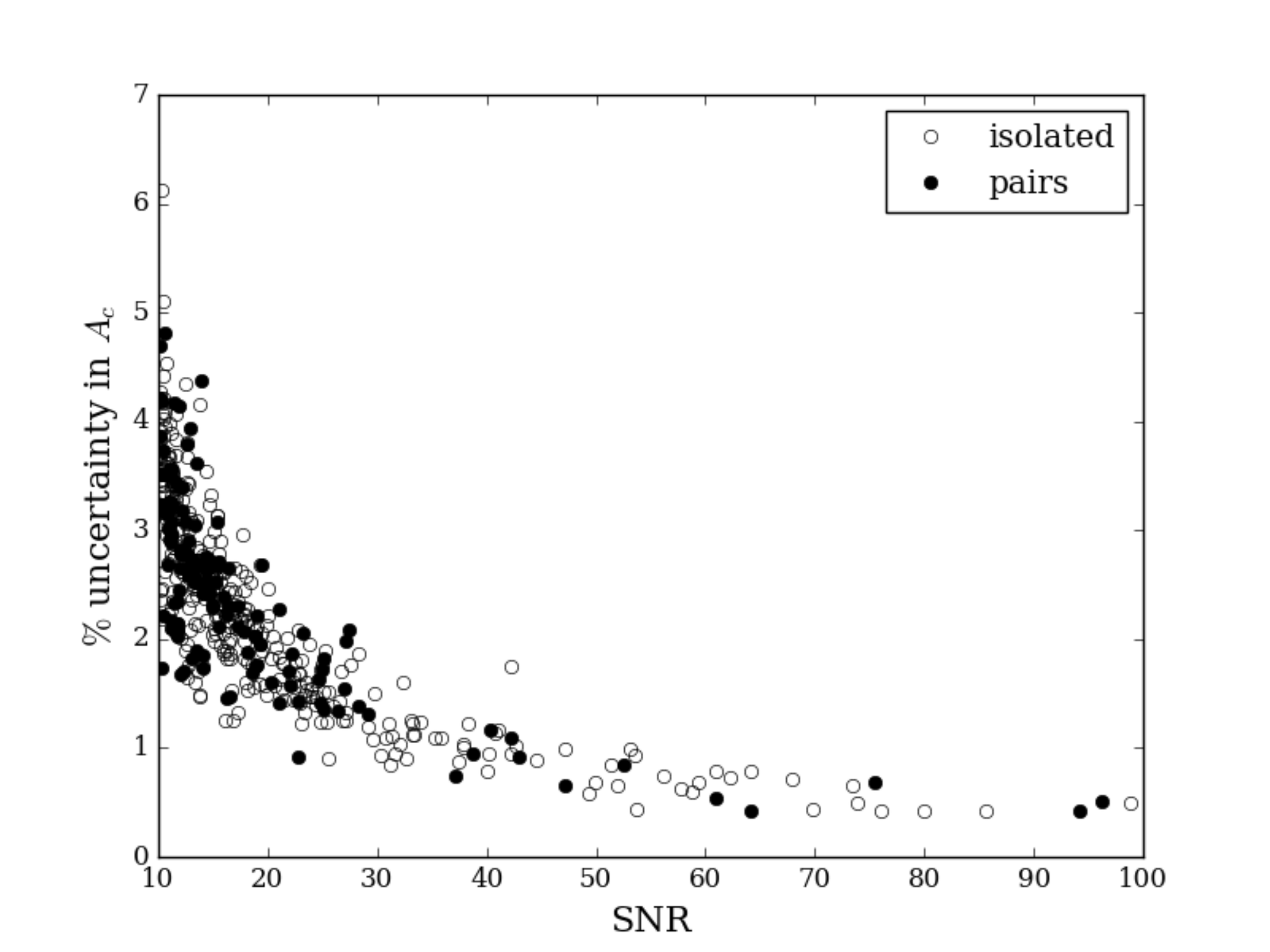}
\caption{Percentage estimated uncertainty on $A_c$ as a function of SNR. Filled circles correspond to the pair sample, while open circles mark the isolated galaxies.}
\label{error}
\end{figure*}

\noindent We use the standard deviation of the 1000 $A_c$ measurements calculated for each profile to serve as the estimated $A_c$ uncertainty, and as can be seen in figure \ref{error}, the calculated uncertainties are for the most part less than 5 percent. Since both our pair and isolated samples are drawn from the same survey, we exclude the potential contribution of $\Delta v_{\rm{mean}}$ and pointing errors from our uncertainty calculation as they should impact both samples equally. \\

\section{Results $\&$ Discussion}\label{results}

\begin{figure*}  
\includegraphics[scale = 0.8]{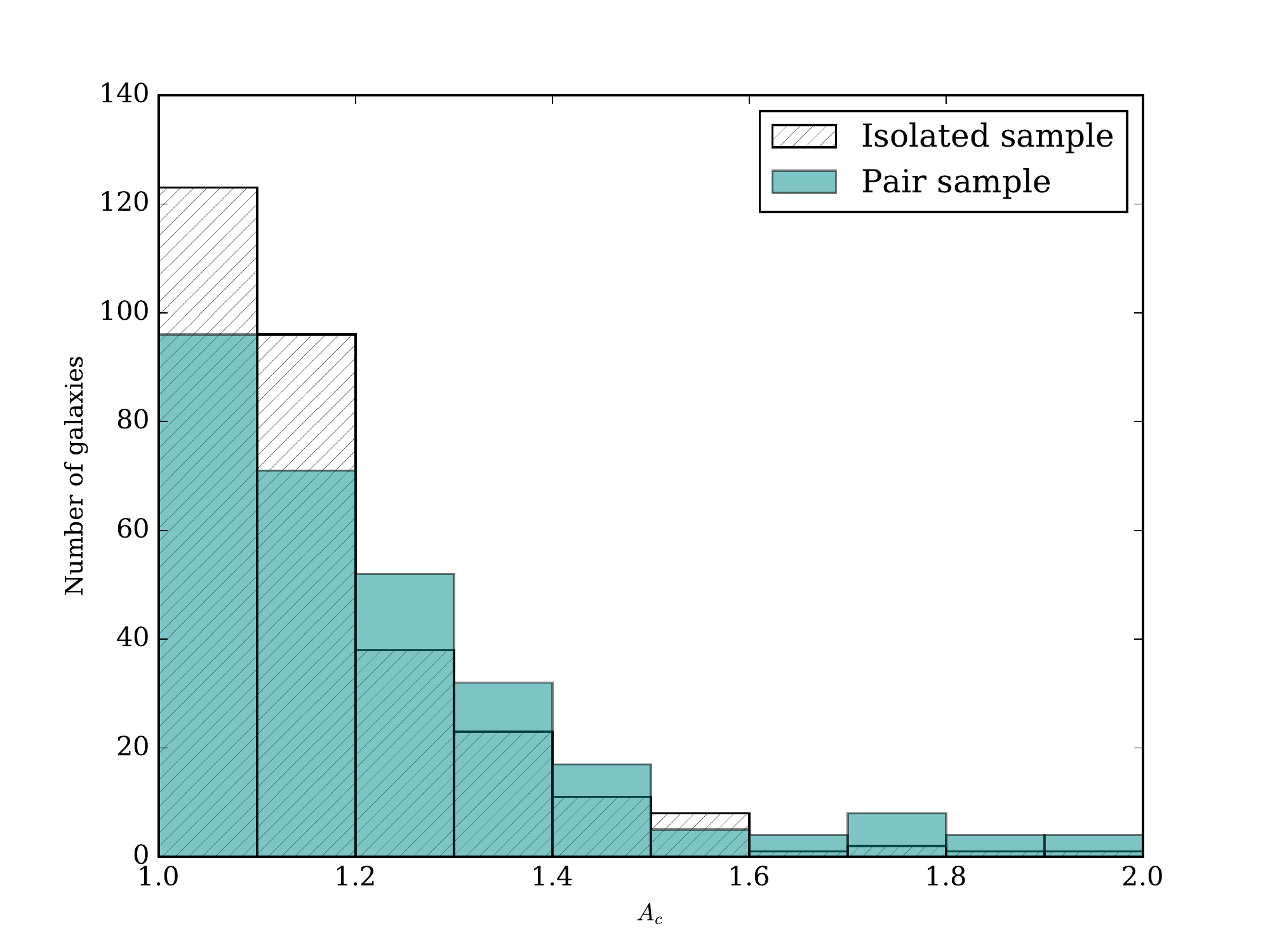}
\caption{Asymmetry distributions for both the pair (dark cyan) and isolated (hatched) galaxy samples.}
\label{result1}
\end{figure*}
\

In figure \ref{result1} we present our results for asymmetry measurements for the close pair sample compared to our isolated sample. The most discernible feature between the two distributions is a longer tail in the pair asymmetry distribution extending towards high asymmetries. %Adopting an asymmetry value of $A_c > 1.26$ (corresponding to the 2$\sigma$ width of \cite{Espada2011}'s asymmetry distribution) to mark the lower limit on high asymmetries, we measure a higher fraction of paired galaxies with high asymmetries compared to isolated galaxies (31 percent vs. 21 percent respectively). 
With the pair and isolated galaxies matched in redshift, and well matched in u-r colour and SNR, we tentatively attribute this enhanced frequency of high asymmetries in the pair sample to environment, and conclude that merger activity is most likely responsible for the measured difference in \hi profile asymmetries between our pair and isolated sample. A statistically significant measure of this difference is provided by the k-sample Anderson-Darling (A-D) test (\cite{Scholz1987}). The k-sample A-D test is a modification of the more widely used Kolmogorov-Smirnov (K-S) test, however that is more sensitive to differences present in the tails of distributions compared to the K-S test, and thus more aptly suited to our data. Bootstrap re-sampling our isolated galaxy sample 10000 times, we measure a mean A-D test statistic between the pair and isolated samples of $A^{2}$ =$12.18$, with a mean p-value = 0.0002, we therefore reject the null hypothesis that they are drawn from the same distribution at the 1percent level.\\
In section \ref{sampleprop} we note that the pair and isolated galaxy samples have different $M_*$ and M$_{HI}$distributions, quantities known to play a role in galaxy evolution. We find no correlation between $A_c$ and both $M_*$ and $M_{HI}$ and thus conclude that these quantities are not responsible for the difference in $A_c$ distributions we measure between the pair and isolated samples. Similarly, we find no correlation between $A_c$ and inclination, where inclination > 30\degr.
\begin{figure*}
\includegraphics[scale=0.7]{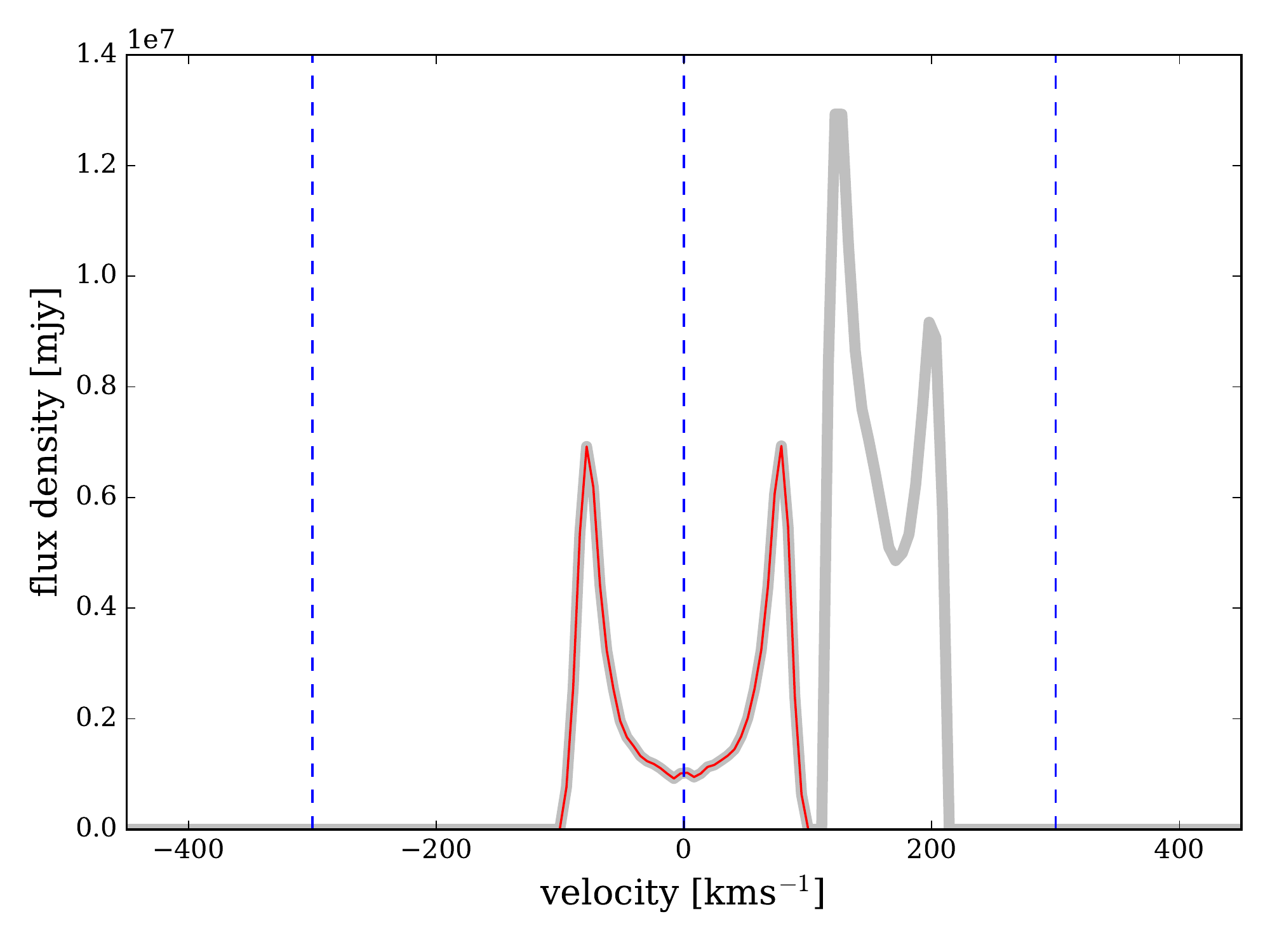}
\caption{The \hi spectrum of a target galaxy extracted from the synthetic \hi data cube is shown in red, with the total extracted \hi mass, including that of nearby neighbour, indicated by the thick grey curve.}
\label{7a}
\end{figure*}
\begin{figure*}
\includegraphics[scale=0.7]{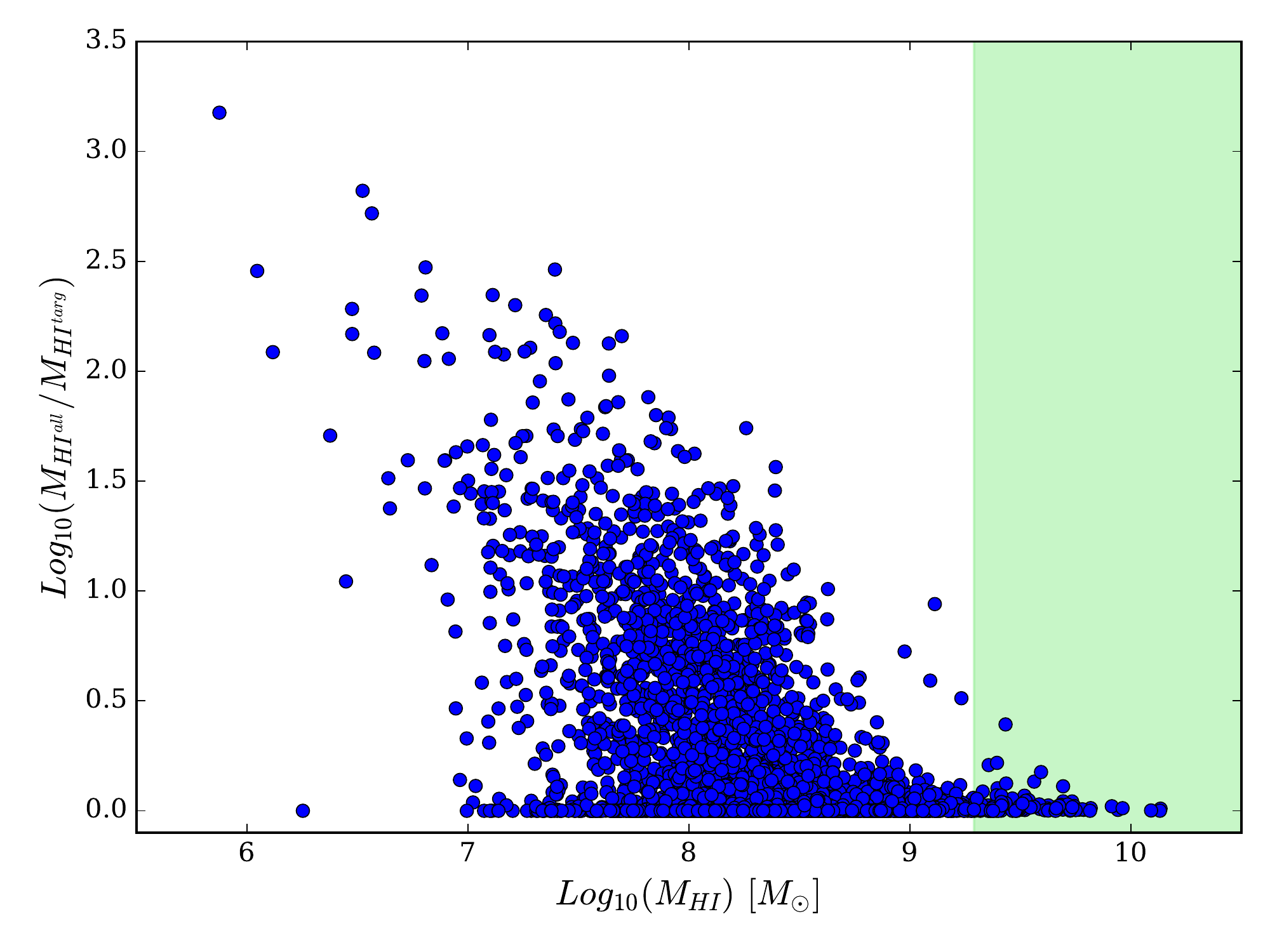}
\caption{Relative amount contaminant flux ($Log_{10}(M_{HI^{all}}/M_{HI^{targ}})$) of the simulated galaxy profiles as a function of \hi mass, with the ALFALFA \hi mass range shaded in green.}
\label{7b}
\end{figure*}
\begin{figure*}
\includegraphics[scale=0.7]{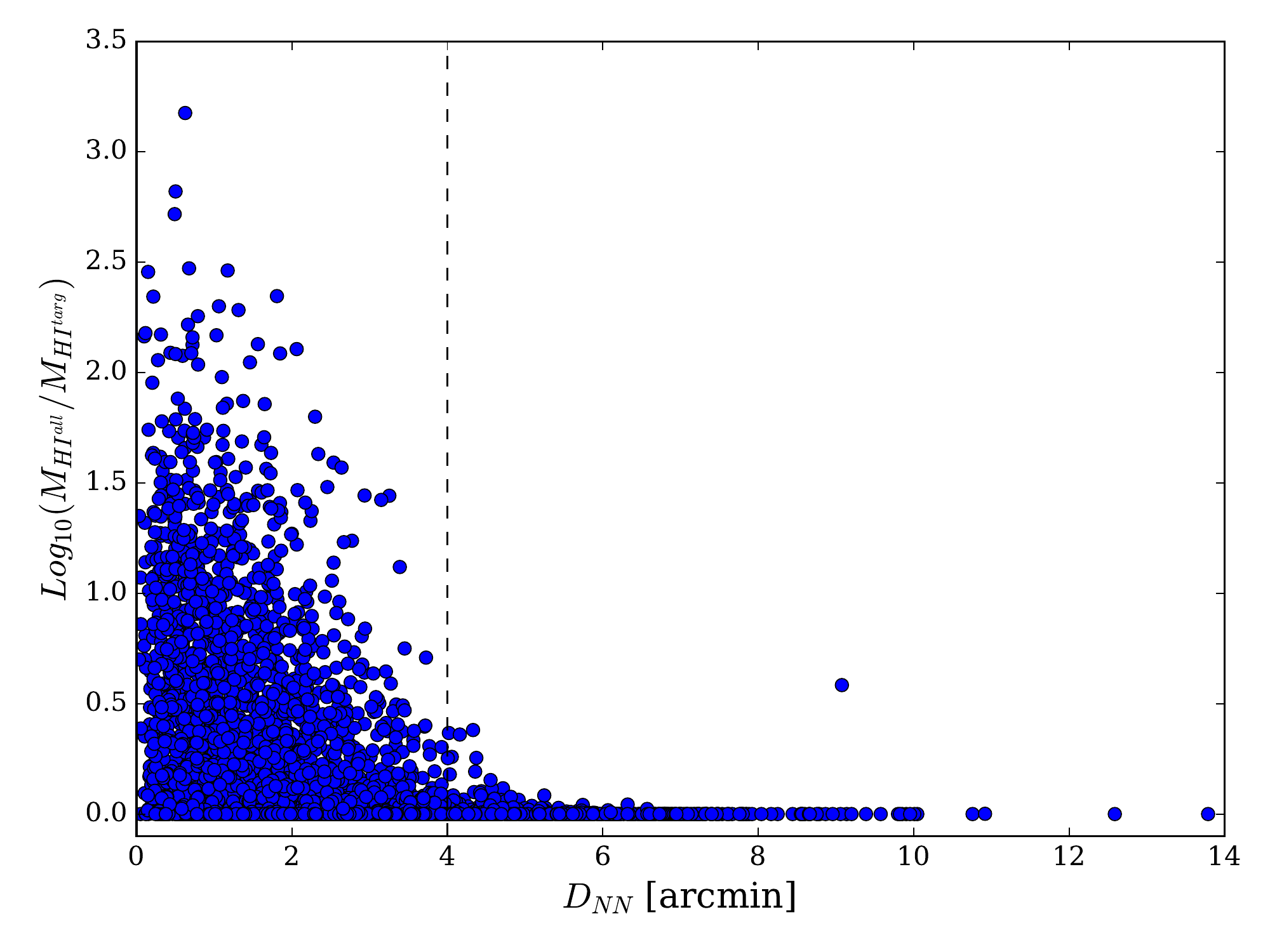}
\caption{Relative amount contaminant flux ($Log_{10}(M_{HI^{all}}/M_{HI^{targ}})$) of the simulated galaxy profiles as a function of angular distance to nearest neighbour (D$_{\rm{NN}}$). }
\label{7c}
\end{figure*}
In order to be able to unambiguously attribute asymmetries in the \hi profiles of our pair sample to merger activity, we need to first ensure that we sufficiently address confusion as the next most likely cause of asymmetry. Our ability to detect and eliminate instances of \hi confusion from our pair sample (as described in \ref{secpairs}) is limited by the ALFAFA resolution ($\sim$3.5'). We cannot account for contamination from \hi sources with flux densities below the ALFALFA noise threshold.  We can, however, make use of simulations to quantify the relative amount of contaminant emission contained in the \hi spectrum of a galaxy.  We make use of a synthetic \hi data cube generated according to the methods presented in \cite{Elson2016}. The cube spans a sky area of 30 square degrees and the redshift range $z<0.06$.  Each of the 3715 galaxies in the cube has an associated set of physical parameters based on the semi-analytic models of \cite{Obreschkow&Meyer2014}. Furthermore, each galaxy has the spatial and spectral distributions of its \hi line emission modelling in a realistic manner.  The entire cube was smoothed to a spatial resolution of 3.5'x3.5' to match the spatial resolution of the ALFALFA data.  The angular size of a spatial pixel in the cube is 30''x30'', whereas the median velocity width of a channel is 5.42 \kms \\
In order to asses the extent to which the \hi spectrum of a galaxy in the cube may be contaminated by emission from neighbouring galaxies, \hi spectra of all 3433 galaxies with $M*>10^{10}$\msol  were extracted using a spatial aperture of 3.5'x 3.5' and a spectral aperture of 230 channels.  The true \hi mass of each galaxy was compared to the total amount of \hi mass spanning 300 \kms either side of its systemic velocity.  This procedure is demonstrated in figure \ref{7a} which shows the \hi spectrum of the target galaxy as the thin red curve and the \hi spectrum of all the extracted mass as the thick grey curve.  In this example, a significant amount of mass from a nearby galaxy falls well within the 600 \kms velocity range centered on the systemic velocity of the target galaxy.  The ratio of the total amount of mass over this range (i.e., the integral of the grey curve) to the mass of the target galaxy (i.e., the integral of the red curve) is M$_{\rm{HI}}$$^{\rm{all}}$/M$_{\rm{HI}}$$^{\rm{targ}}=2.54$.  Figure \ref{7b} shows the logarithm of this ratio as a function of log(M$_{\rm{HI}}$$^{\rm{targ}})$ \msol, where M$_{\rm{HI}}$$^{\rm{targ}}$ is the \hi mass of a target galaxy, for all 3433 of the \hi spectra extracted from the synthetic cube.  Clearly, galaxies with  low \hi masses can contain amounts of contaminant emission that are factors of hundreds to thousands greater than their true \hi mass. However, over the high \hi mass range probed by the ALFALFA data, contamination levels are very low.  This result is further illustrated in figure \ref{7c} which shows the relative amount of contaminant flux as a function of nearest neighbour distance for all \hi spectra extracted from the synthetic cube.  These results give us great confidence that the levels of contamination in the ALFALFA spectra used in this study are negligibly small.
In section \ref{seciso} we discuss how spectroscopic incompleteness of the optical sample could decrease the purity of our isolated galaxy sample, as well as the completeness of our pair sample. We also note that using 2D distance information to locate merger pairs has its limitations in accurately identifying real merger pairs. Ideally we would need 3D position and velocity information to conclusively identify pairs that are going to merge. Without the 3D information we expect our pair sample to be contaminated by false 2D pairs, reducing the purity of our pair sample. Both of these effects would therefore act to increase the similarities between the isolated and pair samples. We therefore expect the measured difference in asymmetries between our pair and isolated samples must be a lower limit on this quantity.
\begin{figure*}
\includegraphics[scale=0.8]{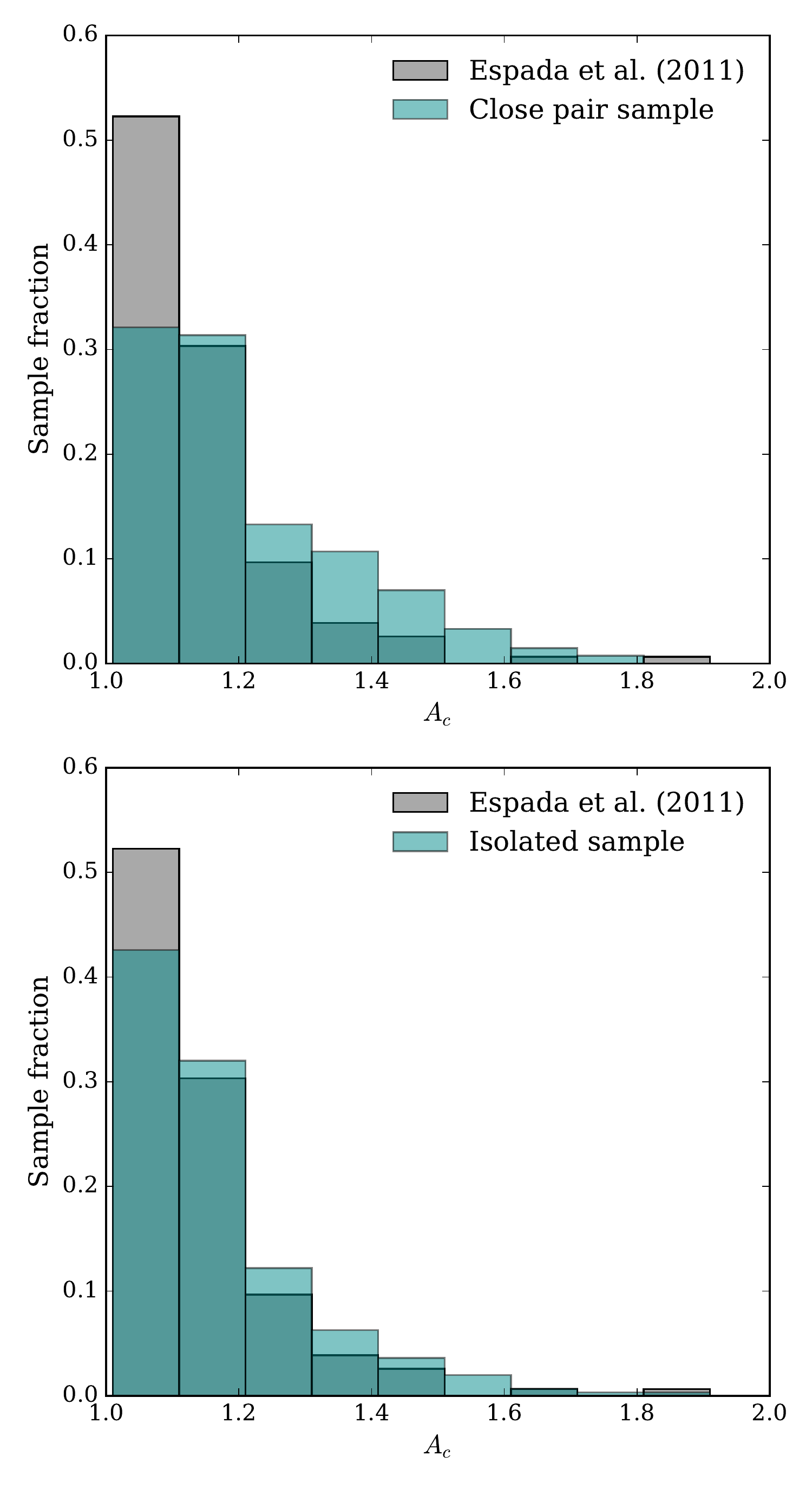}
\caption{Top: Normalized $A_c$ distributions of the \hi refined sample (gray) and our pair sample (dark cyan). Bottom: Normalized $A_c$ distributions of the \hi refined sample (gray) and our isolated sample (dark cyan)}
% * <jamie@ast.uct.ac.za> 2018-12-12T08:52:22.061Z:
%
% ^.
% * <jamie@ast.uct.ac.za> 2018-12-12T08:52:20.264Z:
%
% ^.
\label{espada}
\end{figure*}

\begin{table*}
\caption{Comparison between the \hi asymmetry rate in our pair and isolated galaxies with isolated samples in the literaure.}
\begin{tabular}{ |p{6cm}||p{1cm}|p{1.6cm}|lp{3cm} }
 \hline 
 Galaxy sample & Sample size &   $A_{c}~>1.26 $ & Standard error  \\
 \hline \hline
 \hi refined subsample \citep{Espada2011}   & 166    &  $9\%$ &$2.2\%$ \\
 Haynes et al. (1988)&   104  &  $9\%$& $2.8\%$ \\
 Matthews et al. (1998) & 30 & $17\%$ &$6.8\%$\\
 \hline
 \hi isolated sample (Bok et al. (2017)) & 304& $18\%$& $2.2\%$ \\
 \hi- optical pair sample (Bok et al. (2017)) & 304 & $27\%$& $2.6\%$\\
 \hline \hline

\end{tabular}
\label{table:2}
\end{table*}

Looking to the literature for comparison, \cite{Espada2011} describe the \hi profile asymmetry distribution for their sample of isolated galaxies, the \hi refined sub-sample, as a half Gaussian of width $\sigma = 0.13$. (This sample was selected as having the lowest uncertainties ($\leq5$ percent) in their asymmetry measure.) The 2$\sigma$ level is measured to be at $A_c = 1.26$, above which 9 percent of the sample lies.  \cite{Espada2011} make a direct comparison with the \cite{Haynes1998} isolated sample, showing their asymmetry distribution to also follow a Gaussian of width 0.13, again with 9 percent of the sample having an $A_c$ value $> 1.26$. Making the same direct comparison between our samples and the \hi refined sub-sample, we find that both our isolated and pair samples exhibit higher \hi profile asymmetries (see table \ref{table:2}).  Approximately $ 18$ percent of our isolated profiles have measured $A_c$ values $> 1.26$, while $27$ percent of pairs lie in this asymmetry regime.  We note, however, that while the Gaussian fit is in good agreement with \cite{Espada2011}'s data for the lower asymmetry regime, it does not very accurately recover the high asymmetry tail (see figure 9 in the \cite{Espada2011} paper). Using the A-D test to compare our isolated sample with that of \cite{Espada2011} we do find the samples to be significantly different ($A^2~=~ 5 $, p-value $~= ~0.0005,$). We note, however, that our isolated sample is more similar to the \hi refined sample than our pair sample. We also measure a particularly large difference of $A^{2}$ = $22.18$  (p-value = $0.00002$) between our pair sample and the \hi refined sample. The discrepancy between our isolated sample and those isolated samples in the literature may well be attributed to systematics (different telescopes, resolution, smoothing, sample size). In this regard the \hi refined sub-sample is likely more reliably isolated as the AMIGA team perform a number of follow up observations in different wavebands to more thoroughly ascertain the environment around each galaxy. The \cite{Matthews1998} sample of field galaxies, in comparison to the above-mentioned isolated samples, has $17$ percent of its \hi profiles measuring \hi asymmetries greater than $A_c = 1.26$. While the \cite{Matthews1998} sample is indeed very small, this fraction is more similar to the fraction we measure for our own isolated sample, and suggests that perhaps our isolated sample might be better described as a sample of field galaxies rather than a strictly isolated sample (such as the AMIGA sample). This, however, does not detract from the fact that whether we compare our pair sample asymmetries with our own isolated sample, or with isolated/field samples from the literature, we observe enhanced \hi profile asymmetries for galaxies that are in close pairs. We further argue that due to the potential sensitivity of the asymmetry measurement to systematics, such a comparison is best made when comparative samples are drawn from the same data set. 
We note the possibility that using the ALFALFA heliocentric velocity to mark the center point of our \hi profiles might lead to an underestimate of the profile asymmetry in the case of very asymmetric profiles. ALFALFA define $v_{Helio}$ as the midpoint between the channels at which the flux density drops to 50 percent of each of the two peaks at each side of the spectral feature. If these two velocity values are not placed symmetrically about the true systemic velocity, the derived $v_{Helio}$ is shifted closer to the $v_{50}$ value that is most asymmetric compared to the true systemic velocity- this reduces the quantified profile asymmetry. Since this effect acts to reduce the measured asymmetry, our quantified asymmetries would in the worst cases, be underestimated. However, this should affect profiles in both the isolated and pair samples in the same way. Since our most asymmetric profiles reside in the pair sample, we pose that quantifying this effect would only enhance the difference we measure between the pair and isolated profile asymmetries, and strengthen our result.
Going forward we plan to explore what other mechanisms might be causing \hi profile asymmetries. For a sample of 13 \hi  Magellanic type spiral galaxies, 4 of which have companions, \cite{Wilcots2004} found very little difference in the measured \hi profile asymmetries between the apparently interacting galaxies and non-interacting galaxies. We point out, however, that the \cite{Wilcots2004} sample is very small (only 13 galaxies), and that the optical companions they mention were not spectroscopically confirmed. \cite{Wilcots2004} also find no correlation between optical and \hi asymmetries for their sample. Using optical asymmetries from the \cite{Matthews1998} catalogue of 2D photometric decompositions of the SDSS-DR7 spectroscopic main galaxy sample, we also find no correlation between the optical and HI asymmetries for both our pair and isolated samples. These results suggest that perhaps asymmetries measured in \hi trace merger activity on a different time scale to optical asymmetries. \cite{Espada2011} find for a sample of 166 \hi galaxies that \hi profile asymmetries seem to persist even in the absence of companions. A larger sample is our own isolated galaxy sample (358 galaxies), for which we also measure significant profile asymmetries. These findings suggest an alternative asymmetry driver might be at play. We check the potential dependence of our measured profile asymmetries on the major/minor status of our pair stellar mass ratios, and find a spread of asymmetries for both our major and minor pairs, with only 5 pairs having stellar mass ratios > 2. Outflows and inflows, as well as asymmetric accretion of gas from the cosmic web \citep{Eymeren2008} are also proposed as potential candidates for causing \hi profile asymmetries, for which deeper optical imaging of the sample is required in order to investigate further. In a future paper we will explore the possibility of time scales playing a role in the apparent lack of correlation between optical and \hi asymmetries, as well as alternative \hi profile asymmetry drivers.

We also plan to explore the possibility that our isolated sample is contaminated by real pairs incorrectly identified as isolated galaxies due to having faint companions not detected by SDSS.  With deeper optical imaging we can also investigate the prevalence of faint companions in our isolated sample, and thereby produce a purer isolated sample with which to make the pair/isolated asymmetry comparison. We propose that a purer isolated galaxy sample will only enhance the difference we see in profile asymmetries across the paired and isolated galaxy environments.
Future surveys such as WALLABY\citep{Australian2014}, LADUMA \citep{Holwerda2011}, and the APERTIF shallow and medium-deep surveys \citep{Verheijen2009}, will enable us to study the gas in galaxies in large samples with high spatial resolution, and out to redshifts higher than ever before. At the highest redshifts, however, galaxies will  be unresolved spatially, and it is only the \hi profile with which we will be able to study these early Universe galaxies. The work done here endeavours to maximize the amount of information we can extract from \hi profiles such that when the SKA pathfinder data becomes available, we can begin to characterize the neutral gas content of galaxies over large redshift ranges, and thereby start to put together a more complete and observationally informed picture of galaxy evolution.

\section{Summary}\label{summary}
In summary, a first quantitative look into \hi profile asymmetries in contrasting environments for large samples of paired and isolated galaxies shows that the asymmetry distributions of close pair and isolated galaxies are statistically, and significantly different, with the paired galaxy sample exhibiting an extended asymmetry tail toward higher asymmetries compared to the isolated asymmetry distribution. We see a stronger signal in the asymmetry difference when we compare our pair sample with isolated samples in the literature. 
The work done in this paper suggests that merger activity is responsible for the observed higher frequency of high profile asymmetries in our close pair galaxy sample. We thus put forward that \hi profile asymmetries measured in the high asymmetry regime ($A_c > 1.26$) could be used to infer potential merger activity. While imaging techniques might provide a more robust measure of merger activity, \hi profile asymmetries provide a promising alternative that can already be applied to large samples of galaxies. Furthermore, in the absence of imaging data at high redshifts, employing \hi profile asymmetries as an indicator of merger activity can allow us to estimate merger activity in the early Universe (soon to be probed in \hi by SKA pathfinder telescopes), and test galaxy evolution models. 

\section*{Acknowledgements}
We thank the anonymous referee for their very useful and valuable comments which resulted in a greatly improved paper. We also thank Lourdes Verdes-Montenegro and Michael G. Jones for their very generous expert advice, which not only improved the quality of the paper, but strengthened our results. We thank Martha Haynes for her initial input in defining the project published here, as well as Andrew Baker for his contribution of ideas and guidance throughout the project. JB, SLB, and DGG acknowledge financial support from the National Research Foundation (NRF) of South Africa, JB additionally acknowledges
support from the DST-NRF Professional Development Programme (PDP), the National Astrophysics and Space Science
Programme (NASSP), and the University of Cape Town. 
This work utilizes data from Arecibo Legacy Fast ALFA (ALFALFA) survey data set obtained with the Arecibo L-band Feed Array (ALFA) on the Arecibo 305m telescope. Arecibo Observatory is part of the National Astronomy and Ionosphere Center, which is operated by Cornell University under Cooperative Agreement with the U.S. National Science Foundation. Funding for the SDSS and SDSS-II has been provided by the Alfred P. Sloan Foundation, the Participating Institutions, the National Science Foundation, the U.S. Department of Energy, the National Aeronautics and Space Administration, the Japanese Monbukagakusho, the Max Planck Society, and the Higher Education Funding Council for England. The SDSS Web Site is http://www.sdss.org/. In addition, we make use of data from the Sloan Digital Sky Survey (SDSS DR7). The SDSS is managed by the Astrophysical Research Consortium for the Participating Institutions. The Participating Institutions are the American Museum of Natural History, Astrophysical Institute Potsdam, University of Basel, University of Cambridge, Case Western Reserve University, University of Chicago, Drexel University, Fermilab, the Institute for Advanced Study, the Japan Participation Group, Johns Hopkins University, the Joint Institute for Nuclear Astrophysics, the Kavli Institute for Particle Astrophysics and Cosmology, the Korean Scientist Group, the Chinese Academy of Sciences (LAMOST), Los Alamos National Laboratory, the Max-Planck-Institute for Astronomy (MPIA), the Max-Planck-Institute for Astrophysics (MPA), New Mexico State University, Ohio State University, University of Pittsburgh, University of Portsmouth, Princeton University, the United States Naval Observatory, and the University of Washington.

%The Acknowledgements section is not numbered. Here you can thank helpful colleagues, acknowledge funding agencies, telescopes and facilities used etc. Try to keep it short.

%%%%%%%%%%%%%%%%%%%%%%%%%%%%%%%%%%%%%%%%%%%%%%%%%%


\bibliographystyle{mnras}
\bibliography{correct_paper1}

%\begin{thebibliography}{99}
%\end{thebibliography}

%%%%%%%%%%%%%%%%%%%%%%%%%%%%%%%%%%%%%%%%%%%%%%%%%%

%%%%%%%%%%%%%%%%% APPENDICES %%%%%%%%%%%%%%%%%%%%%

%If you want to present additional material which would interrupt the flow of the main paper,it can be placed in an Appendix which appears after the list of references.

%%%%%%%%%%%%%%%%%%%%%%%%%%%%%%%%%%%%%%%%%%%%%%%%%%

% Don't change these lines
\bsp	% typesetting comment
\label{lastpage}
\end{document}